\documentclass[aps,twocolumn,showpacs,showkeys,groupedaddress]{revtex4}

\usepackage{amssymb}
\usepackage{graphicx}
\usepackage{bm}

\makeatletter
\newcommand\erfc{\mathop{\operator@font erfc}\nolimits}
\def\slashchar#1{\setbox0=\hbox{$#1$}
   \dimen0=\wd0 \setbox1=\hbox{/} \dimen1=\wd1
   \ifdim\dimen0>\dimen1 \rlap{\hbox to \dimen0{\hfil/\hfil}} #1
   \else  \rlap{\hbox to \dimen1{\hfil$#1$\hfil}} / \fi}

\makeatother

\begin{document}

\title{Pion transition form factor and distribution amplitudes in 
large-$N_c$ Regge models}

\author{Enrique \surname{Ruiz Arriola}} \email{earriola@ugr.es}
\affiliation{Departamento de F\'{\i}sica At\'omica, Molecular y
Nuclear, Universidad de Granada, E-18071 Granada, Spain}

\author{Wojciech Broniowski} \email{Wojciech.Broniowski@ifj.edu.pl}
\affiliation{The H. Niewodnicza\'nski
Institute of Nuclear Physics, Polish Academy of Sciences, PL-31342
Krak\'ow, Poland}
\affiliation{Institute of Physics, \'Swi\c{e}tokrzyska Academy,
PL-25406~Kielce, Poland} 

\begin{abstract}
We analyze the $\pi \gamma^\ast \gamma^\ast$ amplitude in the
framework of radial Regge models in the large-$N_c$ limit.  With the
assumption of similarity of the asymptotic Regge $\rho$ and $\omega$
meson spectra we find that the pion distribution amplitude is {\em constant}
in the large-$N_c$ limit at the scale $Q_0$ where the 
QCD radiative corrections are absent -- a result found earlier in a class of
chiral quark models.  We discuss the constraints on the couplings from
the anomaly and from the limit of large photon virtualities, and find
that the coupling of the pion to excited $\rho$ and $\omega$ mesons
must be asymptotically constant. We also discuss the effects of the
QCD evolution on the pion electromagnetic transition form factor.
Finally, we use the Regge model to evaluate the slope of the form factor 
at zero momentum and compare the value to the experiment, finding
very reasonable agreement. 
\end{abstract}

\pacs{12.38.Lg, 11.30, 12.38.-t}
\keywords{pion transition form factor, pion distribution amplitude, 
Regge models, large-$N_c$ limit, quark-hadron duality, non-perturbative QCD}
\maketitle

\section{Introduction} 
\label{sec:intro}

The study of exclusive processes in
QCD~\cite{Lepage:1980fj,Chernyak:1983ej} has been a permanent
challenge both on the theoretical as well as the experimental side.
The production of neutral pions by two virtual photons provides the
simplest process where both perturbative and non-perturbative aspects
of strong interactions can be tested. Indeed, the normalization of the
corresponding from factor for real photons is dictated by the chiral
anomaly. In the opposite limit of large photon virtualities the
amplitude factorizes into power corrections and a soft and scale
dependent distribution amplitude (DA) which gives direct information
on the quark content of the pion at a given scale. Actually, for large
momenta the behavior of the DA is controlled by the perturbative QCD
renormalization group
evolution~\cite{Lepage:1980fj,Dittes:1981aw,Mueller:1994cn} in terms
of Gegenbauer polynomials as implied by the conformal symmetry~(for a
review see, {\em e.g.}, Ref.~\cite{Braun:2003rp}) and the high-$Q^2 $
power expansion.  Then at $Q^2 \to \infty $ one has the fixed point
result, {\em e.g.}, $\varphi^{(2)}(x,Q^2) \to 6 x(1-x)$. For low
scales genuinely non-perturbative evolution can be tackled by
transverse lattice methods~\cite{Dalley:2001gj,Burkardt:2001mf} (for a
review see e.g. Ref.~\cite{Burkardt:2001jg}). The first and second
moments of the DA have been computed on Euclidean
lattices~\cite{DelDebbio:2002mq,DelDebbio:2005bg,Gockeler:2005jz}. The
QCD sum rules have been applied to the leading twist-2
DA~\cite{Belyaev:1997fu}.  Measurements of the transition form factor
have been undertaken by the CELLO~\cite{Behrend:1990sr} and
CLEO~\cite{Gronberg:1997fj} collaborations. An analysis of the lowest
Gegenbauer moments $a_2$ and $a_4$ has been carried out
by~\cite{Schmedding:1999ap}
and~\cite{Bakulev:2001pa,Bakulev:2002uc,Bakulev:2003cs,Bakulev:2005cp}. Higher
twists have been analyzed in the context of light cone sum
rules~\cite{ Agaev:2005rc}.  A direct measurement of the DA has been
presented by the E791 collaboration~\cite{Aitala:2000hb}.  For a
concise review on all these developments see, {\em e.g.},
Ref.~\cite{Bakulev:2004mc} and references therein.

Most calculations dealing with the pion transition form factor and
more specifically with the DA involve {\em quarks as explicit degrees
of freedom}. This appears rather natural but the principle of the
quark-hadron duality suggests that it should also be possible to make
these calculations entirely in terms of the complete set of hadronic
states. In fact, the large $N_c$-limit of
QCD~\cite{'tHooft:1973jz,Witten:1979kh} makes quark-hadron duality
manifest at the expense of introducing an infinite number of weakly
interacting stable mesons and glueballs. The large-$N_c$-limit may in
fact be regarded as a model-independent formulation of the quark
model.  Actually, chiral quark models are particular realizations
implementing in a rather natural way this large-$N_c$ behavior at the
one-quark-loop approximation and several calculations have been made
along these lines~\cite{Petrov:1997ve,Petrov:1998kg,Heinzl:2000ht,
Praszalowicz:2001wy,Anikin:2000bn,Anikin:2000rq,Dorokhov:2002tq,Dorokhov:2002iu,RuizArriola:2002bp,RuizArriola:2003bs}
(for a review see, {\em e.g.},
Ref.~\cite{RuizArriola:2002wr}). Despite the subtleties regarding the
correct implementation of chiral Ward
identities~\cite{RuizArriola:2002wr,Bzdak:2003qe}, chiral quark models
by themselves cannot be ``better'' than the large-$N_c$ limit, as they
form a particular model realization of this limit.  Surprisingly, up
to now there has been remarkable little information on the quark
content of hadrons based solely and directly on the large-$N_c$ limit
and the quark-hadron duality ideas, without resorting to specific
low-energy quark models.

The purpose of this paper is to fill this gap and analyze the pion
transition form factor in the original spirit of the large-$N_c$
limit.  We impose chiral constraints at low energies and QCD short
distance constraints at high energies and extract from there the DA
in a {\em Regge model with infinitely many resonances} where the
radial squared mass spectrum is assumed to be linear. This
near-linearity is supported by the experimental analysis of
Ref.~~\cite{Anisovich:2000kx}. We do the analysis in the absence of
radiative corrections which must be introduced via QCD evolution,
hence our calculation corresponds to some reference scale $Q_0$ where
the radiative QCD corrections are absent.  Remarkably, we obtain the
same main result as found previously by us in chiral quark
models~\cite{RuizArriola:2002bp,RuizArriola:2003bs}, namely, the
leading-twist pion DA is {\em constant at the scale $Q_0$}. Our
calculation provides an explicit example of quark-hadron duality in an
exclusive process.

Up till now the calculations within the framework of large-$N_c$ Regge
models have been mainly restricted to two-point
functions~\cite{Golterman:2001nk,Simonov:2001di,Golterman:2002mi,%
Afonin:2004yb,Afonin:2006da,RuizArriola:2006gq}. On the other hand,
large-$N_c$-motivated calculations of three-point functions with short
distance constraints have been carried out with a {\em finite} number
of resonances, which enabled getting model-independent results for
vector meson
decays~\cite{Moussallam:1994xp,Knecht:2001xc,Beane:2001uj,Beane:2001em,
Ruiz-Femenia:2003hm}. In this regard there are large $N_c$ studies of
both the pion~\cite{Dominguez:2001zu} and the
proton~\cite{Dominguez:2004bx} electromagnetic form factors. It should
also be mentioned that, echoing some older ideas of
Radyushkin~\cite{Radyushkin:1995pj}, the light-cone wave functions
have recently been computed within the holographic approach to QCD
based on the AdS/CFT
correspondence~\cite{Brodsky:2005en,Brodsky:2006uq} or
meromorphization ideas ~\cite{Radyushkin:2006iz}. In these works the
quark-hadron duality is also exploited.

The outline of the paper is as follows. In Sect.~\ref{sec:pda} we
review DAs in the large-$N_c$ context and fix our notation for the
remainder of the paper. In Sect.~\ref{sec:regge} we undertake the
analysis of the large-$N_c$ Regge model and its consequences for the
pion transition form factor and pion DA. Section~\ref{sec:evol} deals
with aspects of the QCD evolution,  which is a crucial
ingredient of analyses of this kind.  For completeness some technical
details of the LO evolution of the non-singlet DA are provided in
Appendix~\ref{sec:gegen}.

\section{parton Distribution Amplitude  and large $N_c$}
\label{sec:pda}

Partonic distribution amplitudes (DAs) are basic properties of bound
states of QCD. They are defined as matrix elements of quark bilinears
between the vacuum and the hadronic state in question. For instance,
the twist-2 DA of $\pi^a$, $\varphi^{(2)}(x)$, is given by
\begin{eqnarray}
&& \langle 0 | \overline{\psi}(z) \gamma_\mu \gamma_5 \frac{\tau^a}{2} 
[z,-z] \psi(-z)| \pi^a(q) \rangle 
\nonumber \\ &=& i f_\pi(q^2) q_\mu 
\int_0^1  dx e^{i u q\cdot z} \varphi^{(2)}(x),  \label{AV} 
\end{eqnarray}
where $u=2x-1$, $z$ is a coordinate along the light cone, $[z,-z]$
denotes the gauge link operator, and $f_\pi(q^2)$ is the pion decay
form factor, with the pion decay constant $f_\pi(0) \equiv f=86$~MeV
in the chiral limit.  The DAs have the support $x\in [0,1]$,
normalization $\int_0^1 dx \, \varphi(x) = 1$, and satisfy the
crossing relation $\varphi(x) = \varphi(1-x)$.  Obviously the
definition (\ref{AV}) requires an identification of quark degrees of
freedom and also specification of a renormalization scale and scheme.

On the other hand, the pion DA is related to the {\em pion
electromagnetic transition form factor} in the process $\gamma^\ast
\gamma^\ast \to \pi^0$, or more generally, to processes with one pion
and two (virtual or real) photons on external legs. This is a physical
matrix element which does not depend on the renormalization scale and
which is more suitable for our purposes. With the outgoing momenta and
polarizations of the photons denoted as $q_1$,$e_1$ and $q_2$, $e_2$
one finds the amplitude
\begin{eqnarray}
\Gamma^{\mu \nu}_{\pi^0  \gamma^\ast \gamma^\ast } (q_1,q_2)  &=& 
\epsilon_{\mu\nu \alpha \beta}e_1^\mu e_2^\nu q_1^\alpha q_2^\beta 
F_{\pi \gamma^\ast \gamma^\ast}(Q^2,A),
\end{eqnarray} 
where the pion transition form factor $F_{\pi \gamma^\ast
\gamma^\ast}$ depends on the total virtuality, $Q^2$, and the photon
asymmetry, $A$,
\begin{eqnarray}
Q^2 = -(q_1^2 + q_2^2 ), \; A = \frac{q_1^2 - q_2^2}{q_1^2+ q_2^2}, \; -1 \le A  \le 1.  
\end{eqnarray} 
Equivalently, $q_1^2 = -\frac{(1+A)}2 Q^2$, $q_2^2=-\frac{(1-A)}2
Q^2$.  For large virtualities one finds the standard twist
decomposition of the pion transition form factor \cite{Lepage:1980fj},
\begin{eqnarray}
F_{\pi^0 \gamma^\ast \gamma^\ast} (Q^2, A ) =  J^{(2)} (A)
\frac1{Q^2} +  J^{(4)} (A) \frac1{Q^4} + \dots , \label{twist}
\end{eqnarray} 
with 
\begin{eqnarray} 
\!\!J^{(2)}(A) &=& \frac{4 f}{N_c } \int_0^1 \!\!\! dx
 \frac{\varphi_\pi^{(2)}(x)}{1-u^2 A^2}, \label{J2} \\ 
\!\!J^{(4)}(A) &=& \frac{8 f \Delta^2 }{N_c} \int_0^1 \!\!\! dx \frac{
 \varphi_\pi^{(4)} (x) [1+u^2 A^2]}
 {\left[1-u^2 A^2\right]^2}, \label{J4}
\end{eqnarray}
involving the subsequent DAs. The above results hold modulo the
logarithmic corrections incorporated by means of the QCD evolution of the
DAs, $\varphi_\pi^{(n)} (x) \to \varphi_\pi^{(n)}
(x,Q)$~\cite{Braun:2003rp}. At infinitely large momentum one reaches
an ultraviolet fixed point behavior \mbox{$\varphi_\pi^{(2)} (x,\infty) = 6
x(1-x)$}, regardless of the value of the DA at some finite scale.  It
is important to realize that, at least in the perturbative regime, QCD
evolution {\it requires} an identification of the powers in a twist
expansion and the corresponding coefficients inherit a logarithmic
momentum dependence, $J^{(n)}(A) \to J^{(n)}(A,Q^2) $ provided their
value is known at some reference scale $Q_0$. Let us recall that power
corrections are a high-energy manifestation of low-energy
non-perturbative phenomena.

\begin{figure}[b]
\includegraphics[angle=0,width=0.18\textwidth]{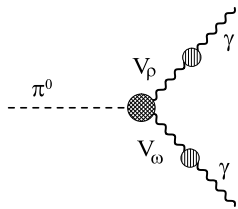}
\caption{The pion electromagnetic transition amplitude in the large-$N_c$ limit.}
\label{fig:diag}
\end{figure}

In the large-$N_c$ limit the vacuum sector of QCD becomes a theory of
infinitely many non-interacting mesons and
glueballs~\cite{'tHooft:1973jz,Witten:1979kh}, hence hadronic
amplitudes may be calculated as tree-level processes, where the
propagators are saturated by infinitely many sharp meson (glueball)
states.  In our case the relevant diagram is shown in
Fig.~\ref{fig:diag}: the pion first couples to a pair of vector mesons
$V_\rho$ and $V_\omega$, which then transform into photons.  From
symmetry constraints, one vector meson has even $G$-parity
($\rho$-type) and the other one odd $G$-parity ($\omega$-type).  Thus,
for a massless pion we have
\begin{eqnarray}
F_{\pi^0 \gamma^\ast \gamma^\ast} (Q^2,A) &=& 
\sum_{V_\rho,V_\omega} \frac{F_{V_\rho}(q_1^2) F_{V_\omega}(q_2^2) 
G_{\pi V_\rho V_\omega}(q_1^2,q_2^2)}{(q_1^2 - M_{V_\rho}^2)(q_2^2 -M_{V_\omega}^2 )}
\nonumber \\ &+& (q_1 \longleftrightarrow q_2), \label{eq:amp}
\end{eqnarray} 
where $F_{V_\rho}$ and $F_{V_\omega}$ are the current-vector meson
couplings and $G_{\pi V_\rho V_\omega}$ is the coupling of two vector
mesons to the pion.  At the soft photon point corresponding to the
neutral pion decay $\pi^0 \to 2 \gamma $ the chiral anomaly matching
condition imposes the normalization
\begin{eqnarray}
F_{\pi^0 \gamma^* \gamma^* } (0,0) &=& \sum_{V_\rho V_\omega} 
\frac{2F_{V_\rho}(0) F_{V_\omega}(0) G_{\pi V_\rho V_\omega}(0,0)}
{M_{V_\rho}^2 M_{V_\omega}^2} \nonumber \\ &=& \frac{1}{4 \pi^2 f}. 
\label{anomc}
\end{eqnarray} 
This consistency constraint, realized in nature, can be always
satisfied in models by an appropriate choice of the couplings.

The large-$Q$ expansion of Eq.~(\ref{eq:amp}) would naively yield pure
power corrections, assuming a {\it finite} number of
resonances~\cite{Moussallam:1994xp,Knecht:2001xc,
Ruiz-Femenia:2003hm}. The situation with infinitely many resonances
requires some care, since the coefficients of the twist expansion
involve positive powers of the meson masses and require 
regularization. We will illustrate below the situation with an
explicit model.
 
Another, more subtle, question regards the role of radiative
corrections in the large-$N_c$ limit. This problem also arises in
chiral quark models which are specific model realizations of the
large-$N_c$ limit with explicit quark degrees of
freedom~\cite{Petrov:1997ve,Petrov:1998kg,Heinzl:2000ht,
Praszalowicz:2001wy,Anikin:2000bn,Dorokhov:2002tq,
RuizArriola:2002bp,RuizArriola:2003bs,RuizArriola:2002wr}. We take
over the viewpoint adopted in our previous
work~\cite{RuizArriola:2002bp,RuizArriola:2003bs,RuizArriola:2002wr},
namely, we consider a situation where all perturbative radiative
corrections are switched off. This corresponds to an identification of
the power corrections at a given reference scale $Q_0$. Thus, our
calculations in the large-$N_c$ limit determine the {\em initial
condition} for the QCD evolution, of the form $\varphi_\pi^{(n)}
(x,Q_0)$. An obvious advantage of such an approach is that {\em after}
the evolution the DAs comply to the known asymptotic QCD behavior.
We discuss this issue in detail in Sect.~\ref{sec:evol}.

\section{Large-$N_c$ Regge Models}
\label{sec:regge}

Now we proceed to the basic analysis of this paper, namely 
a study of Eq.~(\ref{eq:amp}) 
at large $Q^2$. The idea is as follows: we adopt a 
model for the spectra and couplings, calculate the amplitude, and compare to 
Eq.~(\ref{J2},\ref{J4}). 
For the spectra we take the radial Regge model 
\begin{eqnarray}
M^2_{V_\rho}(n) &=& M^2_{V_\omega}(n)= M^2 + a n,
\end{eqnarray} 
which assumes for simplicity the same spectra in the $\rho$ and $\omega$ channels.
is well fulfilled~\cite{Anisovich:2000kx} in the experimentally
explored region.  We also assume that the $M$ and $a$ parameters are
isospin-independent, {\em i.e.}  are the same for the $\rho$- and
$\omega$-type mesons.  We investigate possible departures from this
assumption later on.  We also take constant, {\em
{\em i.e.} $n$-independent} values 
\begin{eqnarray}
F_{V_\rho}&=&F_\rho={\rm const.}, \nonumber \\
F_{V_\omega}&=&F_\omega={\rm const.},
\end{eqnarray} 
as follows from matching of the predictions of the radial Regge model
for the vector correlator to QCD
\cite{RuizArriola:2006gq}.  Standard vector-meson dominance yields 
the relation $F \equiv F_\rho =N_c F_\omega$.  The matching yields
\begin{eqnarray} 
a= 2\pi \sigma = \frac{24\pi^2}{N_c} F^2, \label{c2} 
\end{eqnarray}
where $\sigma$ is the long-distance string tension.
If we use $F = 154~{\rm MeV}$ from the
$\rho \to 2 \pi $ decay~\cite{Ecker:1988te} we get $\sqrt{\sigma} = 546~{\rm MeV}$,
while the lattice calculation of Ref.~\cite{Kaczmarek:2005ui} gives
a similar order-of-magnitude estimate, $\sqrt{\sigma} = 420~{\rm MeV}$. 
Condition (\ref{c2}) originates solely from the asymptotic
spectrum and is insensitive to the low-lying states, whose parameters 
(mass, couplings) may depart form the asymptotic values.

Using the standard Feynman trick we rewrite Eq.~(\ref{eq:amp}) in the
form
\begin{eqnarray}
&& F_{\pi^0 \gamma^\ast \gamma^\ast} (Q,A) = \sum_{n, n'=0}^\infty 
\int_0^1 dx \times \label{nnp}
\\ && \frac{ (F^2 / N_c) G_{n n'}(Q^2,A) }{\left \{
    M^2 +a n x + a n'(1-x) +\frac12 Q^2 [ 1+u A] \right \}^2 }
\nonumber \\ && + (A \longleftrightarrow -A), 
\nonumber
\end{eqnarray} 
where $G_{n n'}$ is the coupling of the pion to the $n$ and $n'$
states belonging to the $\rho$ and $\omega$ 
Regge towers. This coupling may involve
diagonal ($n=n'$) and non-diagonal ($n \neq n'$) terms, hence we
introduce $n'=n+d$. The double sum may then be transformed into
\begin{eqnarray}
\sum_{n,n'=0}^\infty A_{nn'}=\sum_{d=0}^\infty \sum_{m=0}^\infty 
\left ( 1-\frac{\delta_{0 d}}{2} \right ) (A_{m,m+d}+A_{m+d,m}),
\nonumber \\
\end{eqnarray}
where the factor $(1-{\delta_{0 d}}/{2})$ is introduced for the correct
counting the the diagonal and non-diagonal terms.  On purely physical
grounds we expect that the non-diagonal couplings are suppressed,
because in that case the radial wave functions involve different
number of nodes and the overlap is reduced. Thus it is reasonable to
assume that $G_{n,n+d}$ is decreasing sufficiently fast with
increasing $d$ such that Eq.~(\ref{nnp}) makes sense.  Note that
reasonable as it is, this assumption has sound consequences, as it
makes the sum over $n$ and $n'$ in Eq.~(\ref{nnp}) finite. With no
suppression the sum diverges logarithmically, however, any power-law
suppression of $G_{n,n+d}$ with $d$ makes the expression well-behaved.
At asymptotic $Q^2$ and fixed $d$ the coupling $G_{n,n+d}$ might a
priori depend on $n$. We show, however, that this is impossible, as it
would lead to violation of the twist expansion (\ref{twist}).  Indeed,
assume that $G_{n,n+d}= G_{n+d,n} \sim n^{\alpha(d,Q^2,A)}
g(d,Q^2,A)$.  Then, using the Euler-McLaurin summation formula we can
we transform the sum into an integral and get the asymptotic behavior
\begin{eqnarray}
&& F_{\pi^0 \gamma^\ast \gamma^\ast} (Q,A) \sim \sum_{d}
\left ( 1-\frac{\delta_{0 d}}{2} \right )
\int_0^1 dx \int_{n_0}^\infty dn \label{nnp2}
\\ && \left ( \frac{ n^{\alpha(d,Q^2,A)} g(d,Q^2,A)}{\left \{
    M^2 +a n + a d x + \frac12 Q^2 [ 1+u A] \right \}^2 } \right . \nonumber \\
&& \left . 
+ \frac{ n^{\alpha(d,Q^2,A)} g(d,Q^2,A)}{\left \{
    M^2 +a n + a d (1-x) + \frac12 Q^2 [ 1+u A] \right \}^2 } \right ) \nonumber \\
&& \sim  \sum_{d}\left ( 1-\frac{\delta_{0 d}}{2} \right )
 \int_0^1 dx \, g \frac{ [1-u A]^{\alpha-1} 
+ [1+u A]^{\alpha-1} }{ (Q^2)^{1-\alpha}}, 
\nonumber
\end{eqnarray} 
whereas Eqs.~(\ref{twist},\ref{J2}) enforce the leading behavior in
the form $1/[Q^2 (1-u^2 A^2)]$. This implies that the only natural way
to match to QCD (in the absence of radiative corrections) is to assume
that at large $Q^2$ and $n$ we have $\alpha=0$ and $g$ depending only
on $d$.  In other words, {\em the coupling $G_{n,n+d}$ at large $Q^2$
and $n$ is independent of $n$, $Q$, and $A$.} Thus for simplicity we
take $G_{n,n+d}= G_{n+d,n}= g(d)$ for all, even low, $n$.  The
obtained behavior is reminiscent of the asymptotic independence of the
coupling $F_V$ of $Q^2$ and $n$ in the case of the two-point vector
correlator. It is worth stressing that each term in Eq.~(\ref{nnp})
goes as $1/Q^4$ and it is the summation over $n$ which changes the
power to $1/Q^2$, yielding the leading twist DA properly. Any finite
truncation does not do the job. In order to generate the $1/Q^2$
behavior one needs the proper asymptotic density of states, such as in
the Regge model.

With the formula for the polygamma function,
\begin{eqnarray}
\sum_{n=0}^\infty \frac{1}{(u + v n)^2}=\frac{1}{v^2} \psi^{(1)}\left ( \frac{u}{v} \right ),
\end{eqnarray} 
we obtain from Eq.~(\ref{nnp}) the equation
\begin{eqnarray}
&& F_{\pi^0 \gamma^\ast \gamma^\ast} (Q,A) = \frac{F^2}{N_c a^2} \sum_{d}
\left ( 1-\frac{\delta_{0 d}}{2} \right )
 g(d)
\int_0^1 dx \times \label{nnp3} \nonumber \\
&& \left [ \psi^{(1)}\left ( \frac{M^2}{a} + d x +\frac{1}{2a} Q^2 [1+u A] \right )
\right . \label{nx}\\
&& \left . + \psi^{(1)}\left ( \frac{M^2}{a} + d (1-x) +\frac{1}{2a} Q^2 [1+u A] 
\right ) \right ] \nonumber \\ &&+ (A \longleftrightarrow -A).
\nonumber
\end{eqnarray} 
At large $Q^2$ we use the expansion 
\begin{eqnarray}
\psi^{(1)}(A Q^2+B)=\frac{1}{A Q^2}+\left ( \frac{1}{2}-B \right ) \frac{1}{A^2 Q^4}+ \dots
\end{eqnarray}
which yields
\begin{eqnarray}
&& F_{\pi^0 \gamma^\ast \gamma^\ast} (Q,A) = \frac{8F^2}{N_c a} \sum_{d}
\left ( 1-\frac{\delta_{0 d}}{2} \right )
 g(d)
\int_0^1 dx \times \label{nnp4} \\
&& \left ( \frac{1}{Q^2(1-u^2 A^2)} - 
\frac{[2M^2+a(d-1)](1+u^2 A^2)}{Q^4(1-u^2 A^2)^2} +\dots \right ).\nonumber 
\end{eqnarray}
Comparison to Eq.~(\ref{J2},\ref{J4}) gives immediately the identification
\begin{eqnarray}
\varphi^{(2)}(x) = 1, \;\;\; \varphi^{(4)}(x) = 1,  \label{constphi}
\end{eqnarray} 
with the normalization conditions 
\begin{eqnarray}
&&\sum_{d}\left ( 1-\frac{\delta_{0 d}}{2} \right ) g(d)=\frac{12\pi^2 f}{N_c}, 
\label{BLnorm}\\
&&\Delta^2=-\frac{N_c}{24\pi^2 f} \sum_{d}\left ( 1-\frac{\delta_{0 d}}{2} \right ) g(d)
[2M^2+a(d-1)]). \nonumber
\end{eqnarray}
Note that $g(d)\sim N_c^{-1/2}$, as required for 3-meson couplings,
and $\Delta^2 \sim N_c^0$. The sign of $\Delta^2$ is formally not
constrained.  We stress that the above results hold at the scale $Q_0$
at which the radiative gluon corrections are not present.

Let us consider in a greater detail 
the simplest case where no non-diagonal couplings are present, 
$g(d)=G \delta_{d0}$. Then Eqs.~(\ref{BLnorm}) yield 
\begin{eqnarray}
G=\frac{24\pi^2 f}{N_c}, \;\;\; \Delta^2=\frac{a}{2}-M^2. \label{nd0}
\end{eqnarray}
With $M=m_\rho=770$~MeV the parameter $\Delta^2$ is positive when
$\sqrt{\sigma} > 434$~MeV.  At $\sqrt{\sigma} = 500$~MeV we obtain
$\Delta=439$~MeV, which agrees in the order of magnitude with
quark-model estimates \cite{Anikin:1999cx,RuizArriola:2003bs}.

At $Q^2=0$ the anomaly condition (\ref{anomc}) enforces the relation
\begin{eqnarray}
\psi^{(1)}\left ( \frac{m_\rho^2}{2\pi \sigma }\right ) = \frac{\sigma
N_c}{4\pi f^2},
\label{rel}
\end{eqnarray}
which is very well satisfied for $\sigma=(500~{\rm MeV})^2$, when the
above equation becomes $0.377 \simeq 0.377$. If we use the relation
$m_\rho^2=24\pi^2 f^2/N_c$, then Eq.~(\ref{rel}) becomes
$\psi^{(1)}(z)=3/z$, with $z=m_\rho^2/a \simeq 0.385$, which gives numerically the
relation $\sqrt{\sigma} \simeq 0.64m_\rho=495$~MeV. This shows that
extending the assumption of constancy of the the meson coupling $G$
all the way from the asymptotic region down to $Q^2=0$ preserves the
constraints of the model. Moreover, it leads to very reasonable results
and allows to remarkably well determine the string tension from the
meson phenomenology using the chiral anomaly matching condition.

Let us now discuss the effects of modifying the Regge model. Suppose
the masses $M$ and the slope parameters $a$ were different in the
$\rho$ and $\omega$ channels.  If $a_\rho \neq a_\omega$ then the
expansion (\ref{J2},\ref{J4}) is violated. Note, however, that in that
case we would have different asymptotic density of $\rho$ and $\omega$
states, which is not possible in a theory with strict $SU(2)_F$
symmetry. On the other hand, it is possible to split $m_\rho$ and
$m_\omega$.  In that case still $\varphi^{(2)}=1$, but higher-twist
DAs become $x$-dependent. We know from experimental data that the
$\omega$-$\rho$ mass splitting is tiny, so that dependence should not
be strong.  At any case, the result of constant $\varphi^{(2)}$ in the
large-$N_c$ limit (at the scale $Q_0$) seems very robust.

Several comments referring to the interpretation of our results are in
order. We recall that the constancy of pion DAs has been originally
obtained by the present authors in the Nambu--Jona-Lasinio
model~\cite{RuizArriola:2002bp} as well as in the Spectral Quark
Model~\cite{RuizArriola:2003bs}, which are particular realizations of
the large-$N_c$ limit. A key ingredient of these calculation was the
correct implementation of chiral symmetry through the Ward-Takahashi
identities. On the other hand, calculations based on nonlocal quark
models originally obtained bumped
distributions~\cite{Praszalowicz:2001wy}, close in shape to the
asymptotic forms.  Later calculations with more careful implementation
of PCAC resulted in much flatter results~\cite{Bzdak:2003qe}, with the
DA remaining non-zero at the end-points $x=0, 1$. The trend to a flat
distribution can also be seen in transverse lattice calculations at
low transverse scales~\cite{Dalley:2001gj,Burkardt:2001mf}. The
present calculation is founded on more general background, using only
the facts that at large-$N_c$ we deal with a purely mesonic theory and
that the confined meson spectrum may be described by the radial Regge
model.

Another comment refers to the absence of explicit quarks in the
present approach.  Interestingly, the calculation, although referring
to the partonic structure, as seen in Eq.~(\ref{AV}), never explicitly
introduces partons of spin 1/2. The $x$ variable enters from the
Feynman representation of the product of two vector-meson propagators
and is later identified with the Bjorken $x$ by matching to the QCD
expressions (\ref{J2},\ref{J4}). The identification is unique, since
DAs are universal for all $Q^2$ and $A$ and are only functions of $x$
(and the model parameters).

\section{QCD evolution for large and finite $N_c$}
\label{sec:evol}

An important point, not only for our calculation, but for all model
calculations, is the question of the energy scale where the obtained
predictions for the DAs hold. Distribution amplitudes depend on the
scale, while our result corresponds to a fixed reference scale $Q_0$.
The QCD evolution is crucial, eventually evolving the DAs into their
asymptotic forms \mbox{$\varphi_{\rm as}^{(2)}=6x(1-x)$}, \mbox{$\varphi_{\rm
as}^{(4)}=30x^2(1-x)^2$}, \dots.  At leading order the evolution for
the leading-twist component is very simple.  One method is to use the
Gegenbauer moments (see Appendix~\ref{sec:gegen}), which evolve with
the {\em evolution ratio}
$[\alpha(Q^2)/\alpha(Q_0^2)]^{\gamma_n^{(0)}/2 \beta_0}$. Since
$\beta_0=11N_c/3-2N_f/3$, and $\gamma_n^{(0)} \sim (N_c^2-1)/(2 N_c)$
the exponent reaches a finite value in the large-$N_c$ limit. As
expected, the quark contribution (depending on the number of active
flavors $N_F$) is only $1/N_c$ suppressed. For instance, the second
Gegenbauer moment behaves as
\begin{eqnarray}
\frac{a_2 (Q)}{a_2 (Q_0)} =
\left(\frac{\alpha(Q)}{\alpha(Q_0) } \right)^{\frac{75 (N_c^2 -1)}{12 N_c (11 N_c - 2 N_F)}}
\end{eqnarray} 
From the present calculation one gets $a_2 (Q_0) = 7 /18 $. For
$N_F=3$ the exponent changes from $50/81 $ at $N_c=3$ to $25/44$ at
$N_c = \infty$, an effect at the $10\%$ level only. Thus, the
(perturbative) evolution is never switched off. Obviously, our result
for the DA cannot hold at all scales. As we said, it refers to a
particular reference scale $Q_0$.  It is precisely at that situation
where the DAs are constant functions of $x$.  In
order to pass to other scales, the evolution is necessary and its
effect is strong. In particular, the evolution changes the end-point
behavior near $x=0,1$, for instance the twist-2 component
$\varphi^{(2)} \sim x$ near $x=0$ and $\varphi^{(2)} \sim 1-x$ near
$x=1$ (see Refs.~\cite{RuizArriola:2002bp} for details). This effect
is very important, in particular for the analysis of processes with
one real photon, where $A=1$. Then the numerators in integrands of
Eq.~(\ref{J2},\ref{J4}) involve powers of $x(1-x)$ and the transition
form factor is well-behaved only when the end-point behavior of the DA
cancels the singularity. In short, {\em the QCD evolution is mandatory
if we want to compare the model predictions to the data}.

Having said
that, we note that we can compute exactly the unevolved large-$N_c$ pion
transition form factor, which amounts to carrying the $x$ integration in
Eq.~(\ref{nx}). For the diagonal model, $g(d)=G \delta_{d0}$, we find
\begin{eqnarray}
F_{\pi^0 \gamma^\ast \gamma^\ast} (Q,A) &=& \frac{2f}{N_c A Q^2} 
\left [ \psi^{(0)}\left ( \frac{M^2}{a} + \frac{ Q^2 (1+A)}{2a}\right ) 
\right . \nonumber \\ && \left. 
- \psi^{(0)}\left ( \frac{M^2}{a} + \frac{Q^2 (1-A) }{2a} \right ) \right ]. \label{full}
\end{eqnarray} 
\begin{figure}[tb]
\includegraphics[angle=0,width=0.485\textwidth]{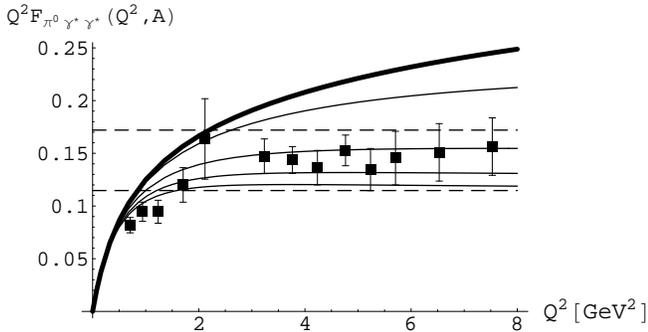}
\vspace{-9mm}
\caption{The pion transition form factor in the large-$N_c$ Regge
model. Solid lines from top to bottom correspond to Eq.~(\ref{full})
at $|A|=1$, $0.95$, $0.75$, $0.5$, and $0$, respectively. The dashed
lines indicate the Brodsky-Lepage limit of $2 f$ for $|A|=1$ (upper curve) and
$4f/3$ for $|A|=0$ (lower curve). The CLEO data points are for $A=1$.}
\label{fig:pex}
\end{figure}
This result is plotted for different values of the asymmetry $A$ in
Fig.~\ref{fig:pex}, where the need for evolution at large $Q^2$ is
vividly seen. The top curve (thick line) is for the un-evolved large-$N_c$
result of Eq.~(\ref{full}) with $A=1$. At large-$Q^2$ we have
\begin{eqnarray}
Q^2 F_{\pi^0 \gamma^\ast \gamma} (Q,|A|=1) &=& \frac{2f}{N_c} \left [
 \log \frac{Q^2}{a}+\psi^{(0)} \left ( \frac{M^2}{a} \right ) \right ] \nonumber \\
&&+ {\cal O}(1/Q^2). 
\end{eqnarray} 
We note the $\log Q^2$ term, whose presence can be traced back to the
singular end-point behavior in the twist expansion
(\ref{twist}-\ref{J4}). As a result, the pure twist expansion is
violated.  The QCD evolution cures this problem, leading to the
asymptotic behavior in accordance to the Brodsky-Lepage limit,
indicated with the upper dashed line in Fig.~\ref{fig:pex}. Therefore
the QCD evolution is necessary to comply to the formal limit, as well
as to the experimental data.  More pictorially, the evolution takes
the tail of the model calculation and brings it down to the upper
dashed line, compatible with the data.  We stress again that the data
should not be directly compared to the un-evolved model results plotted
in the figure, yet we notice that at lower $Q^2$, where the effects of
evolution should be smaller, the comparison is quite reasonable.  At
lower values of $|A|$ the effects of evolution are not as strong as at
$|A|=1$, moreover, the pure twist expansion ({\em i.e.} the expansion
in powers of $1/Q^2$ without logs) holds. In the symmetric limit of
$A=0$ we have
\begin{eqnarray}
F_{\pi^0 \gamma^\ast \gamma^\ast} (Q,A=0) = \frac{2f}{N_c a}
\psi^{(1)}\left ( \frac{2M^2+Q^2}{2a} \right ).
\end{eqnarray} 

\begin{figure}[tb]
\includegraphics[angle=0,width=0.5\textwidth]{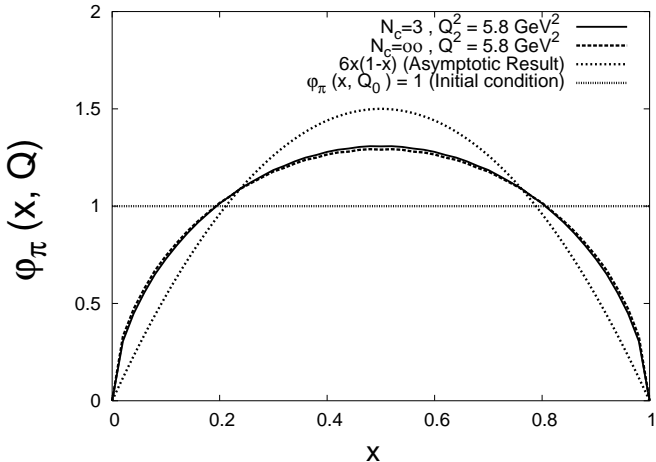}
\vspace{-3mm}
\caption{ The pion distribution amplitude evolved to the scale $Q^2 =
(2.4 {\rm GeV})^2 $ for the cases $N_c=3$ and $N_c=\infty$.  The value
for the evolution ratio $\alpha(Q) / \alpha(Q_0)=0.15$ is based on
the analysis of the CLEO data of Ref.~\cite{Schmedding:1999ap}.  We
also show the un-evolved DA, $\varphi_\pi(x,Q_0)=1$, and the
asymptotic DA, $\varphi_\pi (x,\infty)=6x(1-x)$.}
\label{fig:pda-evol}
\end{figure}

Finally, let us mention that the determination of the reference scale
$Q_0$ can only be done at present using perturbative evolution. In our
previous work~\cite{RuizArriola:2002bp} we used the second Gegenbauer
moment $a_2 =0.12 \pm 0.03 $ at $Q=2.4~{\rm GeV}$ extracted from the
experiment~\cite{Schmedding:1999ap} with the assumption $a_k =0 $ for
$k > 2$.  This allowed us to make the LO estimate for the
evolution ratio $\alpha(Q) / \alpha(Q_0)=0.15 \pm 0.06 $. In
Fig.~\ref{fig:pda-evol} we show the corresponding LO Gegenbauer
evolution for $N_c=3 $ and $N_c= \infty$ to the scale $Q^2 = 5.8~{\rm
GeV}^2$. As we see the $N_c$-effect is tiny and is in fact comparable
with the uncertainties induced by the evolution
ratio~\cite{RuizArriola:2002bp}. Let us note that despite the
perturbative nature of our evolution the similarity of these evolved
results to non-perturbative transverse lattice calculations with a
transverse lattice size of about $a_\perp \sim 0.5-0.7~{\rm fm}$ is
indeed striking~\footnote{This could be understood if a conversion
factor to $\overline{\rm MS}$ scale of $\mu = 2 \pi / a_\perp $ would
be taken, yielding $Q^2 \sim 3.2-6.3~{\rm GeV}^2$}. The LO reference
scale of our estimate turns out to be $Q_0 =0.320 \pm 0.045~{\rm GeV}
$ (for $\Lambda_{\rm QCD}= 0.224~{\rm GeV}$) -- a rather low value
which suggests the usage of higher-order
evolution~\cite{Mueller:1994cn} or even non-perturbative
evolution~\cite{Dalley:2001gj,Burkardt:2001mf}.  Our constant DA
evolved at LO to the CLEO scale $Q^2 = 5.8~{\rm GeV}^2$ yields the
value $Q^2 F_{\gamma^*, \pi \gamma} (Q) / (2 f_\pi) = 1.25 \pm 0.10 $
~\cite{RuizArriola:2002bp}, higher but compatible within two standard deviations
to the experimental CLEO value of $0.83 \pm 0.12$. This actually
speaks in favor of small NLO corrections, however work on
higher-order evolution should definitely be pursued in the future. In
addition, the higher twist corrections should also be included in such
an analysis, which is nontrivial.

Expansion of Eq.~(\ref{full}) at low-$Q^2$ yields
\begin{eqnarray}
F_{\pi^0 \gamma^\ast \gamma^\ast} (Q,A) &=& 
\frac{2f}{3a} \psi^{(1)} \left ( \frac{M^2}{a}\right ) +
\frac{f Q^2}{3a^2} \psi^{(2)} \left ( \frac{M^2}{a}\right ) \nonumber \\ 
&+& \frac{f (A^2+3)Q^4}{36a^3} \psi^{(3)} \left ( \frac{M^2}{a}\right ) +
\dots \label{lq} 
\end{eqnarray} 
Note that the $Q^2$ term is independent of the asymmetry $A$, as is also
apparent from Fig.~\ref{fig:pex}. The corresponding slope reads 
\begin{eqnarray} 
b_\pi &=& -\left[\frac1{F_{\pi^0 \gamma \gamma^\ast} (Q,0)}
\frac{d}{d Q^2} F_{\pi^0 \gamma \gamma^\ast} (Q,0)
\right]\Big|_{Q^2=0} \nonumber \\ &=& -\frac1{2 a}
\frac{\psi^{(2)} \left ( \frac{M^2}{a}\right )} {\psi^{(1)} \left (
\frac{M^2}{a}\right )}
\end{eqnarray}
Numerically, taking $M = m_\rho=770$~MeV we get the value $b_\pi = 1.39~{\rm
GeV}^{-2}$ for $\sigma=(400~{\rm MeV})^2$, $b_\pi = 1.51~{\rm
GeV}^{-2}$ for $\sigma=(500~{\rm MeV})^2$, and $b_\pi = 1.58~{\rm
GeV}^{-2}$ for $\sigma=(600~{\rm MeV})^2$.  These Regge model
estimates are in very reasonable agreement with the experimental
values quoted in the PDG~\cite{PDBook}: $b_\pi = (1.79 \pm 0.14 \pm
14) {\rm GeV}^{-2}$ originally reported by the CELLO
collaboration~\cite{Behrend:1990sr}, obtained from an extrapolation
from high-$Q^2$ data to low $Q^2$ by means of generalized vector meson
dominance, $b_\pi = (1.4 \pm 1.3 \pm 2.6) {\rm GeV}^{-2}$ given in
\cite{Farzanpay:1992pz}, and $b_\pi = (1.4 \pm 0.8 \pm 1.4) {\rm
GeV}^{-2}$ given in \cite{MeijerDrees:1992qb}.

Finally, let us comment on the recent findings within the holographic
approach~\cite{Brodsky:2005en,Brodsky:2006uq} where the light-cone
wave functions have been computed appealing to the AdS/CFT
correspondence and the meromorpization approach of
Ref.~\cite{Radyushkin:2006iz}. The holographic wave-functions
correspond to a spectrum which behaves linearly in the mass, $M(n)
\sim n$.  Based on a conformal-based mapping Brodsky and
Teramond~\cite{Brodsky:2005en} get the asymptotic DA $
\varphi_\pi^{(2)}(x) = 6 x (1-x) $ while the corresponding Parton
Distribution Function (PDF) is $f_\pi (x)= 6x (1-x) $. If, instead, a
twist-based mapping is considered~\cite{Brodsky:2006uq} these authors
get $\varphi_\pi^{(2)}(x) = (8 /\pi) \sqrt{x(1-x)} $ and a PDF $f_\pi
(x) = 1$.  According to the meromorphization approach of
Radyushkin~\cite{Radyushkin:2006iz} one gets $\varphi_\pi^{(2)}(x) =
(8 /\pi) \sqrt{x(1-x)} $ and $f_\pi (x) =1 $ for scalar quarks and $
\varphi_\pi^{(2)}(x) = 6 x (1-x)$ , the asymptotic DA, and $f_\pi (x)
= 6 x(1-x)$ for spin 1/2 quarks. This latter situation is exactly the
kind of situation that was found in the NJL quantized on the light
cone~\cite{Heinzl:2000ht}. An asymptotic PDA suggests a reference
scale $Q_0 = \infty$ or the lack of LO evolution of $\phi^2$, which
assumes the asymptotic form at all scales. On the other hand the
corresponding PDF yields a $100\%$ momentum fraction carried by the
quarks, while for $Q_0=\infty$ one expects that all momentum fraction
is carried by the gluons. This poses an interpretation difficulty for
these approaches which should be cleared out.

\section{Conclusion}
\label{sec:conc}

We summarize our main points.  The radial Regge model predictions for
the $\pi^0 \gamma^\ast \gamma^\ast$ amplitude can be matched to the
QCD twist expansion in the absence of radiative corrections in the
large-$N_c$ limit. The matching requires that the coupling of the pion
to a pair of Regge $\rho$ and $\omega$ mesons, $G_{\pi V V'}$, is
constant for highly excited Regge states and large momenta. The
twist-2 pion distribution amplitude is then found to be constant in
the $x$-variable, conforming to the earlier predictions made in a
class of chiral quark models. Thus, we provide an explicit example of
quark-hadron duality for an exclusive process. Moreover, higher-twist
DAs may or may not be constant, depending on the details of the Regge
model. If $m_\rho=m_\omega$, then the twist-4 DA is also constant.
The Regge model couplings are constrained by matching the chiral
anomaly for real photons and the high-momentum behavior, {\em cf.}
Eqs.~(\ref{BLnorm},\ref{nd0}) for highly virtual photons. If we
further assume a model with constant diagonal couplings, a consistency
relation (\ref{rel}) involving the string tension is found, which is
well supported by the data. As a result the string tension is found to
scale with the square of the $\rho$-meson mass. The Regge model
results is very reasonable predictions for the slope of the pion
electromagnetic transition form factor at zero momentum. The estimates
depend weakly on the string tension, and compare quite well to the
existent measurements. Finally, we have noted that similarly to the
case of chiral quark models, the QCD evolution is necessary in the
present Regge model to achieve the correct large-momentum behavior of
the pion transition form factor.

\begin{acknowledgments}

Useful correspondence with Anatoly~V.~Radyushkin, Guy de Teramond and
Stan Brodsky is gratefully acknowledged.  This research is supported
by the Polish Ministry of Education and Science, grants 2P03B~02828
and 2P03B~05925, by the Spanish Ministerio de Asuntos Exteriores and
the Polish Ministry of Education and Science, project 4990/R04/05, by
the Spanish DGI and FEDER funds with grant no. FIS2005-00810, Junta de
Andaluc\'{\i}a grant No. FQM-225, and EU RTN Contract CT2002-0311
(EURIDICE).

\end{acknowledgments}

\appendix

\section{LO evolution of DA}
\label{sec:gegen} 

The LO-evolved distribution amplitude reads~\cite{Lepage:1980fj,Mueller:1994cn}
\begin{eqnarray}
\varphi_{\pi}^{(2)}(x,Q) &=& 6x(1-x){\sum_{n=0}^\infty}'  C_n^{3/2} ( 2 x -1)
a_n (Q),
\label{eq:evolpda} 
\end{eqnarray}
where the prime indicates summation over even values of $n$ only. The matrix
elements, $a_n(Q)$, are the Gegenbauer moments given by
\begin{eqnarray}
a_n (Q)&=& \frac23 \frac{2n+3}{(n+1)(n+2)} \left(
 \frac{\alpha(Q_{})}{\alpha(Q_0) } \right)^{\gamma_n^{(0)} / (2
 \beta_0)} \times \nonumber \\ &&\int_0^1 dx C_n^{3/2} ( 2x -1)
 \varphi_{\pi}^{(2)} (x ,Q_0),
\label{Geg}
\end{eqnarray}
with $C_n^{3/2}$ denoting the Gegenbauer polynomials, and the
non-singlet anomalous dimension reads 
\begin{eqnarray}
\gamma_n^{(0)} &=& -2 C_F \left[ 3 + \frac{2}{(n+1)(n+2)}- 4
\sum_{k=1}^{n+1} \frac1k \right], \nonumber \\ 
\beta_0 &=& \frac{11}3 C_A -
\frac23 N_F,
\end{eqnarray}
with $C_A = N_c $, $C_F = (N_c^2 - 1)/(2N_c)$, and $N_F$ being the
number of active flavors, which we take equal to three. The LO running
is then $\alpha(Q) = (4 \pi / \beta_0)/ \log (Q^2 / \Lambda_{\rm
QCD}^2)$.  Taking as initial condition
\begin{eqnarray}
\varphi_{\pi}^{(2)} (x,Q_0) = 1 \, , 
\label{start}
\end{eqnarray}
one we gets immediately
\begin{eqnarray}
a_n (Q)&=& \frac23 \frac{2n+3}{(n+1)(n+2)} \left(
\frac{\alpha(Q_{})}{\alpha(Q_0) } \right)^{\gamma_n^{(0)} / (2
\beta_0)} \, . 
\label{OurGeg}
\end{eqnarray}


\end{document}